----------
X-Sun-Data-Type: default
X-Sun-Data-Description: default
X-Sun-Data-Name: reflec1
X-Sun-Charset: us-ascii
X-Sun-Content-Lines: 847

\documentstyle[twoside,fleqn,espcrc2]{article}

\newcommand{\AmS}{{\protect\the\textfont2
 A\kern-.1667em\lower.5ex\hbox{M}\kern-.125emS}}
\begin{document}
\title{Reflections on the Strong CP Problem}
\author{R. D. Peccei\address{Department of Physics and Astronomy \\
        UCLA, Los Angeles, CA 90095-1547}
\thanks{This work is supported in part by the Department of Energy under
        Grant No. DE-FG03-91ER40662, Task C.}}
         
\begin{abstract}
I discuss how anomalies affect classical symmetries and how, in turn, the
non-trivial nature of the gauge theory vacuum makes these quantum corrections
troublesome.  Although no solution seems in sight for the cosmological
constant problem, I examine three possible approaches to the strong CP
problem involving vacuum dynamics, an additional chiral symmetry, and the
possibility of spontaneous CP or P breaking.  All of these ``solutions" have
their own problems and suggest that, at a deep level, we do not understand
the nature of CP violation.  Nevertheless, it remains
extremely important to search for experimental signals predicted by these theoretical ``solutions", like invisible axions.
\end{abstract}
\maketitle

\section{EFFECTIVE THEORIES AND THEIR FAILURES}

The strong CP problem\cite{RDP} is intimately connected with the failure of
symmetries to survive quantum effects.  Let me illustrate this point by a
simple example.  At the classical level, in general, complex mass terms for
fermions are not necessarily a signal of time reversal (T) violation.  In
fact, offending complex terms in the Lagrangian
\begin{equation}
{\cal{L}}_{\rm mass} = -me^{i\theta}\bar\psi_{\rm L}\psi_{\rm R}
-me^{-i\theta}\bar\psi_{\rm R}\psi_{\rm L}
\end{equation}
can be rotated away provided that the fermion fields have chiral invariant
interactions, such as those provided by gauge interactions.  In this case, one may
perform a chiral rotation
\begin{equation}
\psi_{\rm L}\to e^{i\theta/2}\psi_{\rm L}~; ~~
\psi_{\rm R} \to e^{-i\theta/2}\psi_{\rm R}
\end{equation}
which eliminates the phase $\theta$ from ${\cal{L}}_{\rm mass}$ altogether.

This pleasant situation, however, changes at the quantum level as a result
of the existence of chiral anomalies.\cite{ABJ}  Even though the transformation
(2) is classically allowed (i.e. it does not alter the rest of the
Lagrangian, besides ${\cal{L}}_{\rm mass}$), because the chiral current
is {\bf not} divergenceless at the quantum level, the transformation (2) induces
an equivalent T-violating term.  Specifically, if the field $\psi$
interacts with some non-Abelian gauge field $A^\mu_a$, so that the chiral
current
\begin{equation}
J_5^\mu = \bar\psi\gamma^\mu\gamma_5\psi
\end{equation}
has an anomaly\cite{ABJ}
\begin{equation}
\partial_\mu J_5^\mu = \frac{g^2}{32\pi^2} F_a^{\mu\nu}
\tilde F_{a\mu\nu}~,
\end{equation}
then the transformation (2) changes ${\cal{L}}_{\rm mass}$ to
\begin{equation}
{\cal{L}}_{\rm mass}\to -m(\bar\psi_{\rm L}\psi_{\rm R} +
\bar\psi_{\rm R}\psi_{\rm L}) + \frac{\theta g^2}{32\pi^2}
F_a^{\mu\nu}\tilde F_{a\mu\nu}~.
\end{equation}
The $\theta F\tilde F$ term is C-conserving, but both P- and T-odd.  So, at the
quantum level, complex fermion mass terms are indeed signals of T-violation.

The Standard Model has two such classical symmetry failures.  The first of
these, which is due to the chiral anomaly, is at the root of the strong
CP problem.  Because of the chiral anomaly connected with the gluon field
strength, $G_a^{\mu\nu}$, the full Lagrangian of the Standard Model is augmented
by the following effective interaction\footnote{There is no equivalent
$\theta_{\rm weak}W\tilde W$ interaction involving the $SU(2)$ gauge fields,
because the electroweak interactions possess an overall chiral 
symmetry.\cite{Anselm}}
\begin{equation}
{\cal{L}}^{\rm eff} = \bar\theta\frac{\alpha_3}{8\pi}
G_a^{\mu\nu} \tilde G_{a\mu\nu}~.
\end{equation}
Here $\alpha_3$ is, essentially, the square of the $SU(3)$ coupling constant
$[\alpha_3 = g_3^2/4\pi]$ and $\bar\theta$ is a parameter containing both
QCD and electroweak information:
\begin{equation}
\bar\theta = \theta + \hbox{Arg det}~M~.
\end{equation}
In the above $\theta$ is the QCD vacuum angle, while $M$ is the quark mass
matrix which obtains after the breakdown of the electroweak symmetry.

The second manifestation of the violation of a classical symmetry in the
Standard Model arises in the trace of the energy momentum tensor.  This trace
has an additional piece beyond the ``classical" trace $\theta^\mu_{~\mu}$,
reflecting an anomaly in the dilatational current.  One has, retaining only
the QCD piece of the trace anomaly,\cite{Adler}
\begin{equation}
T^\mu_{~\mu} = \frac{\beta(g_3)}{2g_3} G_a^{\mu\nu}G_{a\mu\nu} +
\theta^\mu_{~\mu}~,
\end{equation}
where $\beta(g_3)$ is the $\beta$-function for QCD.  

What makes the quantum corrections (6) and (8) important is the nontrivial
nature of the gauge vacuum.  I begin by considering the $\theta G\tilde G$
term.  Here, at first sight, it actually seems that this quantum correction
may be ineffective, since the density $G\tilde G$ is a total derivative\cite{Bardeen}:
\begin{equation}
G_a^{\mu\nu}\tilde G_{a\mu\nu} = \partial_\mu K^\mu
\end{equation}
with
\begin{equation}
K^\mu = \epsilon^{\mu\alpha\beta\gamma} A_{a\alpha}
\left\{G_{a\beta\gamma} - \frac{g_3}{3}
f_{abc}A_{b\beta}A_{c\gamma}\right\}~.
\end{equation}
However, the non-trivial nature of the gauge theory vacuum does not allow
one to throw out this total divergence.  It turns out that amplitudes in
which the vacuum gauge field configurations at $t=\pm\infty$ differ by a,
so-called, large gauge transformation\cite{CDG} are associated with
non-trivial $G\tilde G$ configurations.  In fact, the difference in the
indices\footnote{For a pure gauge
field, the index $n$ details how the gauge transformation goes to unity at
spatial infinity: $\Omega_n(\vec r)\to e^{2\pi in}$.\cite{Crewther}} 
of the pure gauge fields at $t=\pm\infty$ is given
by\cite{CDG}
\begin{equation}
\nu = n_+ - n_- = \frac{\alpha_3}{8\pi} \int d^4x
G_a^{\mu\nu}\tilde G_{a\mu\nu}~.
\end{equation}
So, as long as amplitudes with $\nu\not= 0$ are important, one cannot ignore
the quantum correction (6).

There is separate evidence that $\nu\not= 0$ configurations are important in
QCD, related to the apparent non-existence of an approximate chiral
$U(1)_A$ symmetry of this theory.  Since $m_u$ and $m_d$ are much smaller than
the dynamical QCD scale, $\Lambda_{\rm QCD}$, the QCD Lagrangian should
possess an approximate $U(2)_A$ chiral symmetry.  This symmetry is
spontaneously broken by the formation of quark condensates
\begin{equation}
\langle\bar uu\rangle = \langle \bar dd\rangle\sim \Lambda^3_{\rm QCD}~,
\end{equation}
so one expects 4 near Nambu-Goldstone bosons.  Although the pions behave as
such, the $\eta$ meson appears quite different\cite{Weinberg} and, in
fact, $m^2_\eta\gg m^2_\pi$.  This can be understood if the $U(1)_A$ subgroup of $U(2)_A$ is not an (approximate) symmetry at all.  Because of the chiral anomaly,\cite{ABJ}
even in the limit when $m_{u,d}\to 0$, one finds a violation of the
associated $U(1)_A$ chiral charge, with
\begin{eqnarray}
\Delta Q_5 &=& \int d^4x\partial_\mu J_5^\mu \nonumber \\ 
&=& \frac{\alpha_3}{4\pi}
\int d^4x G_a^{\mu\nu}\tilde G_{a\mu\nu} = 2\nu~.    
\end{eqnarray}
So, to understand why the $\eta$ is {\bf not} a Nambu-Goldstone boson it must be
that $\nu\not= 0$ configurations are important in QCD.  Whence, it follows
that also the quantum corrections (6) must be significant, unless for some
reason the parameter $\bar\theta$ is very small or vanishes.

Not only does the QCD vacuum allow the formation of the quark condensates
[cf Eq. (12)] which break chirality, it also permits the formation of a
gluon condensate
\begin{equation}
\langle G_a^{\mu\nu} G_{a\mu\nu}\rangle \sim
\Lambda^4_{\rm QCD}~.
\end{equation}
This latter condensate, in view of Eq. (8), produces a significant vacuum
energy density
\begin{equation}
\langle T^\mu_{~\mu}\rangle = \frac{\beta(g_3)}{2g_3}
\langle G_a^{\mu\nu}G_{a\mu\nu}\rangle +
\langle\theta^\mu_{~\mu}\rangle~.
\end{equation}
One does not really know what the value of $\langle\theta^\mu_{~\mu}\rangle$
is.  Naively, it should be at least of order (100 GeV)$^4$ due to the
breakdown of the electroweak theory.  However, it is possible that
$\langle\theta^\mu_{~\mu}\rangle$ vanishes.  At any rate, using the value
for $\langle G_a^{\mu\nu}G_{a\mu\nu}\rangle$ deduced from QCD sum rules,\cite{sum} the first term in Eq. (15) already produces a vacuum energy
density of order
\begin{equation}
\langle T^\mu_{~\mu}\rangle\sim (350~{\rm MeV})^4~.
\end{equation}
This value is about 45 orders of magnitude {\bf larger} than the
nominal bound for the cosmological constant.\footnote{This bound equates
the cosmological constant to the critical density of the Universe.}
\begin{equation}
\Lambda \leq (3\times 10^{-3}~{\rm eV})^4~.
\end{equation}
Because $\Lambda$ measures the vacuum energy density, one expects that
$\Lambda = \langle T^\mu_\mu\rangle$.  Obviously, the discrepancy between
Eqs. (16) and (17) tells us that this is not the case. The reason for this flagrant violation of our intuition is a present day mystery.

A similar, but slightly less severe, puzzle is presented by Eq. (6).  The
$\bar\theta G\tilde G$ term, because it violates both P and T, can give rise
to a sizeable electron dipole moment for the neutron, unless the angle
parameter $\bar\theta$ is very small.  To calculate the size of this dipole
moment, it is useful to perform a chiral rotation which transform the
$\bar\theta G\tilde G$ term into a complex quark mass term\cite{RDP}
\begin{equation}
{\cal{L}}_{\rm CP-viol.} = i\bar\theta m_q
\left[\bar u \frac{\gamma_5}{2} u + \bar d\frac{\gamma_5}{2} d\right]~.
\end{equation}
One can use the above effective Lagrangian directly to calculate the neutron
dipole moment via the equation
\begin{eqnarray}
d_n\bar n \sigma_{\mu\nu}k^\nu\gamma_5 n &=&
\langle n|T(J_\mu^{\rm em} \nonumber \\
&\times& i\int d^4x {\cal{L}}_{\rm CP-viol.})|n\rangle~.
\end{eqnarray}
To arrive at a result for $d_n$, one insert a complete set of states
$|X\rangle$ in the matrix element above and tries to estimate which set of
states dominates.  In the literature there are two calculations along
these lines.  Baluni\cite{Baluni} uses for $|X\rangle$ the odd parity
$|N^-_{1/2}\rangle$ states which are coupled to the neutron by
${\cal{L}}_{\rm CP-viol.}$.  Crewther, {\it et al.},\cite{CDVW} instead, do
a soft pion calculation where, effectively, $|X\rangle\sim |N\pi_{\rm soft}\rangle$.  The result of these calculations are rather similar and
lead to an expression for $d_n$ whose form could have been guessed at.  
Namely,
\begin{equation}
d_n\sim \frac{e}{M_n} \left(\frac{m_q}{M_n}\right)\bar\theta \sim
\left\{
\begin{array}{l}
2.7\times 10^{-16}~\bar\theta ~~\cite{Baluni} \\
5.2\times 10^{-16}~\bar\theta ~~\cite{CDVW}
\end{array} 
\right.
\end{equation}
The present bound on $d_n$\cite{PDG} at the 95\% C.L., is
\begin{equation}
d_n < 1.1\times 10^{-25}~{\rm e~cm}~.
\end{equation}
Whence, to avoid contradiction with experiment, the parameter $\bar\theta$
must be less than $2\times 10^{-10}$.  Why this parameter, which is the sum
of two disparate terms, $\theta~{\hbox{and Arg det}} M$, should be so small is another mystery. 

\section{APPROACHES TO THE STRONG CP PROBLEM}

There are no believable mechanisms to guarantee that the cosmological constant
either vanishes or satisfies the bound (17).  Obviously, the vacuum energy
density induced by the gluon condensate and other VEVs from spontaneous
symmetry breakdowns apparently either cancel among each other---something
that is difficult to believe, given the different scales involved---or,
somehow, do not end up by contributing to the cosmological constant.  In this
respect, the situation concerning the strong CP problem is better.  Here, at
least, there are some ideas on how perhaps to resolve the conundrum
raised by the presence of the $\bar\theta G\tilde G$ term.

There are three distinct approaches to the strong CP problem.  The first of
these supposes that the vacuum dynamics itself selects $\bar\theta$ to be
zero, leading to no CP-violating effects.  The second imposes
an additional chiral symmetry on the theory\cite{PQ} which dynamically
drives $\bar\theta\to 0$.  The third approach supposes that CP (or perhaps P)
is spontaneously broken, with the resulting theory producing naturally very
small values for $\bar\theta,~\bar\theta \leq 10^{-10}$.\cite{spontaneous}
All three approaches leave a host of questions unanswered and are, in some
sense, unsatisfactory.  However, they do have some experimental
consequences and, indeed, experiments may give us a hint of which of these
approaches may be ultimately viable.  
In what follows, I want to briefly discuss and review
these ``solutions" to the strong CP problem, focusing particularly on their
more troublesome features.

\subsection{Vacuum Dynamics}

There have been various attempts to solve the strong CP problem within
QCD. Although I do not believe that the solution of this problem is
to be found in this direction, let me mention three such possibilities that
have been raised at various times:
\begin{description}
\item{(i)} One knows that the vacuum energy is periodic in $\bar\theta$\cite{Coleman}
\begin{equation}
E_{\rm vac}(\bar\theta) \sim (1-\cos\bar\theta)~.
\end{equation}
Thus, if one were to assume that the correct theory has minimum vacuum
energy, then $\bar\theta=0$ would naturally ensue.  Unfortunately, I know of
no physical principle that demands that one should minimize $E_{\rm vac}$.
\item{(ii)} A more interesting suggestion, perhaps, has been put forth by
Schierholz.\cite{Schierholz}  He argues that it is possible that QCD may not 
confine for $\bar\theta\not= 0$.  Hence, since all indications are that QCD
confines, it must be that $\bar\theta=0$.  Schierholz indeed finds evidence
for a deconfining phase transition at finite vacuum angle in the
CP$^N$ model.  However, it is really difficult to extrapolate from this
result to QCD.  In fact, it is unclear to me what role, if any, the vacuum
angle, or $\bar\theta$, plays for confinement in QCD.  So I do not see how
confinement could force $\bar\theta\to 0$.
\item{(iii)} Finally, there have also been suggestions that the $\theta$
vacuum is an artifact of the boundary condition imposed on the gauge
transformation matrices at spatial infinity.\cite{CDG}  If one does not
impose such boundary conditions, the necessity for the $\theta$-vacuum
disappears and so does the strong CP problem.  However, then one is left again
to understand why the $\eta$ meson does not behave like a Nambu-Goldstone
boson.  For this reason,  
I do not believe that the $\theta$-vacuum is an artifact.
\end{description}

\subsection{The Chiral Solution to the Strong CP Problem}

Because a chiral transformation can change the vacuum angle\cite{JR}
\begin{equation}
e^{+i\alpha\tilde{Q_5}}|\bar\theta\rangle = |\bar\theta + \alpha\rangle~,
\end{equation}
a natural solution to the strong CP problem assumes the existence of some
{\bf additional} chiral symmetry in the Standard Model.  Two suggestions
have been put forth:
\begin{description}
\item{(i)} The lightest quark, the $u$-quark, actually has zero mass,
$m_u=0$.\cite{massless}
\item{(ii)} The Standard Model is invariant under an additional $U(1)$ chiral
symmetry, $U(1)_{\rm PQ}$.\cite{PQ}
\end{description}
The first of these possibilities is disfavored theoretically by a careful analysis of the low energy spectrum of QCD, which is inconsistent with having
$m_u= 0$.\cite{Leutwyler} In addition, in my view, by 
appealing to this ``solution", one has just exchanged one problem for another.  What is the
origin of the chiral symmetry which makes det $M=0$?

I am, of course, prejudiced in favor of the second chiral solution!
Imposing an additional, spontaneously broken, chiral symmetry
$U(1)_{\rm PQ}$ on the Standard Model replaces the static CP-violating
parameter $\bar\theta$ by the dynamical CP-conserving field associated with
the $U(1)_{\rm PQ}$ pseudo Nambu-Goldstone boson---the axion.\cite{WW}
This replacement
\begin{equation}
\bar\theta\to \frac{a(x)}{f}
\end{equation}
introduces into the theory a new parameter $f$, the scale of the spontaneous
breakdown of $U(1)_{\rm PQ}$.

The axion field translates under a $U(1)_{\rm PQ}$ transformation
\begin{equation}
a(x)\stackrel{\rm PQ}{\longrightarrow} a(x) + \alpha f~.
\end{equation}
Thus, in the effective low energy Lagrangian this field will always appear
derivatively coupled, with the exception of a term needed to reproduce the
chiral anomaly in the $U(1)_{\rm PQ}$ current.  Assuming the existence of an
extra $U(1)_{\rm PQ}$ symmetry, one finds\cite{RDP}
\begin{eqnarray}
{\cal{L}}_{\rm low~energy} = {\cal{L}}_{\rm SM} &-&
\frac{1}{2} \partial_\mu a\partial^\mu a + {\cal{L}}_{\rm int}
[\psi;~\frac{\partial^\mu a}{f}] \nonumber \\
&+& [\bar\theta + \frac{a}{f}] \frac{\alpha_3}{8\pi}
G_a^{\mu\nu}\tilde G_{a\mu\nu}~.
\end{eqnarray}
The last term in (26) insures that the $U(1)_{\rm PQ}$ current indeed has the
expected chiral anomaly
\begin{equation}
\partial_\mu J^\mu_{\rm PQ} = \frac{\alpha_3}{8\pi}
G_a^{\mu\nu}\tilde G_{a\mu\nu}~.
\end{equation}
This term effectively also gives the axion field a non-trivial potential.
The minimum of this effective potential
\begin{equation}
0 =\left.\left\langle\frac{\partial V_{\rm eff}}{\partial a}\right\rangle
\right|_{\langle a\rangle} = -\frac{\alpha_3}{8\pi f} \left.
\langle G_a^{\mu\nu}\tilde G_{a\mu\nu}\rangle\right|_{\langle a\rangle}
\end{equation}
occurs at
\begin{equation}
\langle a\rangle = -\bar\theta f~,
\end{equation}
due to periodicity of the $\langle G\tilde G\rangle$ vacuum expectation
value in the relevant $\theta$-parameter: $\bar\theta + (\langle a\rangle/f)$.\cite{PQ}  Obviously, Eq. (29) solves the strong CP problem,
since the coupling of the 
physical axion field $a_{\rm phys} = a-\langle a\rangle$ 
to $G\tilde G$ removes the $\bar\theta G\tilde G$ term.  Furthermore,
the second derivative of the effective potential $V_{\rm eff}$ gives a
small mass for the axion, of order $\Lambda^2_{\rm QCD}/f$:
\begin{eqnarray}
m^2_a &=& \left.\left\langle\frac{\partial^2 V_{\rm eff}}{\partial a^2}
\right\rangle\right|_{\langle a\rangle} = -\frac{\alpha_3}{8\pi f}
\frac{\partial}{\partial a}\left. \langle G_a^{\mu\nu}\tilde G_{a\mu\nu}\rangle\right|_{\langle a\rangle} \nonumber \\ 
&\sim&
\frac{\Lambda^4_{\rm QCD}}{f^2}~.
\end{eqnarray}

The above mechanism works for any value of the parameter $f$ associated with
the scale of $U(1)_{\rm PQ}$ breakdown.  Because all interactions of the
axion scale as $1/f$, as does its mass, the larger this scale is the more
weakly coupled and lighter the axion is.\footnote{The axion is not stable
since, through the electromagnetic anomaly, it can always decay into two
photons.  This lifetime scales as $\tau(a\to 2\gamma)\sim f^5$, so the
axion becomes very long-lived for large values of $f$.}  In our original paper,
Helen Quinn and I chose quite naturally for the $U(1)_{\rm PQ}$ breaking scale
$f$, the
weak scale $v=(\sqrt{2}~G_F)^{-1/2}$.    Unfortunately, weak scale axions,
with $f\simeq v$, are ruled out experimentally,\cite{RDP} so our specific
suggestion is no longer tenable.  However, models where $f \gg v$ are still
perfectly consistent.

If $f\gg v$, it is clear that what triggers the breaking of the
$U(1)_{\rm PQ}$ symmetry must be the vacuum expectation value of some
$SU(2)\times U(1)$ singlet object $\sigma$, with $\langle\sigma\rangle = f$.
Remarkably, astrophysics \cite{abounds} and cosmology \cite{cbounds} put non-trivial constraints on $f$
or, equivalently, the axion mass
\begin{equation}
m_a \simeq 6\left[\frac{10^6~{\rm GeV}}{f}\right] {\rm eV}~.
\end{equation}
These bounds, taken at face value, restrict $f$ to a rather narrow
range
\begin{equation}
5\times 10^9~{\rm GeV} \leq f \leq 10^{12}~{\rm GeV}~.
\end{equation}
Although the above provides interesting phenomenological constraints
for $f$, the real
issue is what physics causes the $U(1)_{\rm PQ}$ symmetry to break down
precisely in the range of scales indicated by Eq. (32).

In superstring models, where axions arise naturally as fundamental fields,\cite{Witten} the natural scale for $f$ is somewhat above the bound
(32).  Typically\cite{Choi}
\begin{equation}
f\simeq \frac{M_{\rm P}}{16\pi^2} \sim 10^{16}-10^{17}~{\rm GeV}~.
\end{equation}
How to reconcile this potentially natural scale for the $U(1)_{\rm PQ}$
breakdown with the bounds of Eq. (32) has led to a number of possible  explanations.
Typically, in these suggestions one removes the incompatibility between Eqs. (32) and (33)
either by complicating the physics or by changing the relevant cosmology.

One way to get $f$ into the observable range (32) is to associate the
$U(1)_{\rm PQ}$ breakdown with some physical intermediate scale.  A simple
example is provided by some recent work of Murayama, Suzuki and Yanagida.\cite{MSY}  These authors achieve their goal by assuming that the
$U(1)_{\rm PQ}$ symmetry results as a radiative effect in a supergravity
theory with a flat potential.  The relevant effective potential has a
(negative) squared mass term for the singlet field $\sigma$ which is 
radiatively generated and of order $m_{3/2}\sim M_W$.  $U(1)_{\rm PQ}$
breaking occurs as a competition of this term with some gravitational
induced non-renormalizable interactions for the singlet field $\sigma$.
The effective potential 
\begin{equation}
V = -m_{3/2}^2|\sigma|^2 + \frac{\lambda|\sigma|^6}{M_{\rm P}^2}
\end{equation}
give a $U(1)_{\rm PQ}$ breaking VEV
\begin{equation}
\langle\sigma\rangle = f\sim (m_{3/2} M_{\rm P})^{1/2} \sim 10^{10}~
{\rm GeV}
\end{equation}
in the needed range of Eq. (32).

Alternatively, one can alter the cosmology, thereby allowing larger values for
$f$.  The upper bound on $f$, due to cosmology, more properly is a bound on
the product of $f$ and the square of an initial misallignment angle
$\bar\theta_i$:\cite{RDP}
\begin{equation}
f\bar\theta_i^2 < 10^{12}~GeV.
\end{equation}
The usual bound follows by assuming, rather naturally that
$\bar\theta_i\sim O(1)$.  However, Linde\cite{Linde} has suggested that
inflationary models (and the anthropic principle) may well prefer initial
misallignment angles $\bar\theta_i \ll 1$.  In this case, $f$ values like
those in Eq. (33) may well be allowed.  Alternatively,\cite{CCK} one can
again raise the bound on $f$ by arranging for a period of large entropy
production for $T \sim \Lambda_{\rm QCD}$.  This effectively reduces the importance of axion oscillations to the energy density of the Universe and
allows for values of $f > 10^{12}~{\rm GeV}$, even if $\bar\theta_{i}
\sim O(1)$.  Of course, relaxing the cosmological bound on $f$ makes the
observability of invisible axions, as the possible source of the dark matter in the Universe,
questionable.  Higher $f$'s implies smaller axion masses and hence
lower frequencies to detect halo axions in resonance cavity
experiments, as well as a smaller signal since this signal scales as
$1/f^2$.\cite{Sikivie}

There is a second troublesome aspect of the $U(1)_{\rm PQ}$ solution to the
strong CP problem, connected with gravitational effects.  One can make
arguments which suggest that gravitational interactions do not allow exact
global symmetries (like $U(1)_{\rm PQ}$) to exist.  Perhaps the simplest
way to understand why this may be so is through the ``No Hair" theorem for
black holes.  Basically, this theorem\cite{Banks} asserts that black holes are
characterized only by a few fundamental quantities, like mass and spin, but
possess otherwise no other quantum numbers.  Because black holes can absorb
particles which carry global charge, while carrying no global charge themselves,
it appears that through these processes one can get an explicit violation
of whatever symmetry is associated with the global charge.  That is, global
charge can be lost when particles carrying this charge are swallowed by a black
hole.

One can parametrize the effect of the breaking of global symmetries by
gravitational interactions by adding to the low-energy Lagrangian
non-renormalizable terms, scaled by inverse powers of the Planck mass
$M_{\rm P}$.  These terms, of course, should be
constructed so as to explicitly violate the symmetries in question, in our case
$U(1)_{\rm PQ}$.  Schematically, therefore, the full Lagrangian of the theory,
besides containing the usual Standard Model terms, should also include some
effective non-renormalizable interactions containing various operators
$O_n$, breaking explicitly $U(1)_{\rm PQ}$
\begin{equation}
{\cal{L}}^{\rm eff}_{\rm grav.~int.} =  \sum_n
\frac{1}{M_{\rm P}^n} O_n~.
\end{equation}
Here the dimension of the operators $O_n$ is $n+4$.  

The addition of the non-renormalizable interactions (37) has a significant
effect and, in general, may vitiate the $U(1)_{\rm PQ}$ solution to the
strong CP problem. \cite{grav}  Even though the interaction terms are scaled by inverse
powers of the Planck mass, these terms both give an additional
contribution to the axion mass and alter the QCD potential,
so that $\bar\theta$ does not finally adjust to zero! 

One can understand what
is going on schematically by sketching the form of the effective axion
potential in the absence and in the presence of the $U(1)_{\rm PQ}$ breaking
gravitational interactions.\cite{BS}  Without gravity, a useful
parametrization for the physical axion effective potential, which follows from
examining the contributions of instantons,\cite{PQ} is
\begin{equation}
V_{\rm axion} = -\Lambda^4_{\rm QCD} \cos a_{\rm phys}/f~.
\end{equation}
This potential displays the necessary periodicity in $a_{\rm phys}/f$, has
a minimum at $\langle a_{\rm phys}\rangle = \bar\theta_{\rm eff} = 0$,
and leads to an axion mass $m_a = \Lambda^2_{\rm QCD}/f$.

Including gravitational effects changes the above potential by adding a sequence of terms involving operators of different dimensions.  Let us just consider
one such term and examine the potential\cite{BS}
\begin{eqnarray}
\tilde V_{\rm axion} = &-&\Lambda^4_{\rm QCD} \cos\frac{a_{\rm phys}}{f}
\nonumber \\
&-&\frac{cf^d}{M_{\rm P}^{d-4}}\cos\left[\frac{a_{phys}}{f} + \delta\right]~.
\end{eqnarray}
Here $c$ is some dimensionless constant and $\delta$ is a CP-violating phase
which enters through the gravitational interactions.  This potential modifies
the formula for the axion mass, giving now
\begin{equation}
m_a^2 \simeq \frac{\Lambda^4_{\rm QCD}}{f^2} + c\frac{f^{d-2}}
{M_{\rm P}^{d-4}}~.
\end{equation}
For $f$ in the range of interest for invisible axions, the second term above
coming from the gravitational effects dominates the QCD mass estimate for the
axion, unless $c$ is extraordinarily small and/or the dimension $d$ is rather
large.  More troublesome still, $\tilde V_{\rm axion}$ now no longer has a
minimum at $\langle a_{\rm phys}\rangle = 0$.  Rather one finds a minimum of
$\tilde V_{\rm axion}$ for values of
\begin{equation}
\bar\theta_{\rm eff} = \frac{\langle a_{\rm phys}\rangle}{f} \simeq
c\sin\delta \frac{f^d}{M_{\rm P}^{d-4}\Lambda^4_{\rm QCD}}~.
\end{equation}
That is, the gravitational effects (provided there is a CP violating phase
associated with them) induce a non-zero $\bar\theta$, even in the presence of
a $U(1)_{\rm PQ}$ symmetry!  To satisfy the bound $\bar\theta\leq 10^{-10}$
again necessitates that $d$ be large and/or that the constant $c$ be
extraordinarily small.\footnote{Note that for $d=5$ the induced term is
enormous since $f^5/M_{\rm P}\Lambda^4_{\rm QCD}\sim 10^{46}$.}

To date there is no clear resolution to this problem and it could be that
these considerations actually destroy the chiral solution to the strong CP
problem.  Because one does not really understand quantum gravity, one cannot
be totally sure of the validity of the above arguments.  Nevertheless, if one  takes these arguments seriously, 
it is gratifying that various loopholes have emerged which preserve the
$U(1)_{\rm PQ}$ solution to the strong CP problem.

The simplest way to avoid any gravitational troubles is to arrange things in
the theory, usually through the imposition of some discrete symmetries, \cite{discrete} so that gravity breaks $U(1)_{\rm PQ}$ only through
high dimension operators.  If $d$ is sufficiently large, such that
\begin{equation}
\frac{f^d}{M_{\rm P}^{d-4}\Lambda^4_{\rm QCD}} < 10^{-10}~,
\end{equation}
then there is no strong CP problem. If (42) holds, furthermore, it turns out that also the gravitational corrections to the axion mass
are negligible and $m_a \sim \Lambda^2_{\rm QCD}/f$.

The second way to avoid problems is if, indeed, the strength of the
non-renormalizable interactions, $c$, is extremely small.  This apparently is
possible in some string theories which have a large compactification radius,\cite{radius} where one can obtain parameters $c < 10^{-56}$!  In that
case, again, the gravitational correction to $\bar\theta < 10^{-10}$ and
the changes to the axion mass are totally irrelevant.  Of course, it gives
one pause to imagine that the understanding of why $\bar\theta < 10^{-10}$
should be through a global symmetry, $U(1)_{\rm PQ}$, whose violation by
gravitational interactions are under control because of the presence of an
{\bf even smaller} parameter $c,~c < 10^{-56}$!

A third possibility, which is perhaps the most interesting from my point of
view, is that there is {\bf no CP-violating phase} in the effective interaction
(39) (i.e. $\sin\delta = 0$).  That is, the gravitationally induced terms 
that violate $U(1)_{\rm PQ}$ do {\bf not} also violate CP.  This is quite an
interesting possibility phenomenologically,\cite{BS} because
$\bar\theta_{\rm eff} = 0$, but the interrelation between the axion mass
$m_a$ and the scale of $U(1)_{PQ}$ breaking, $f$, is changed.  Now, in addition
to a term proportional to $1/f$ the axion mass gets a direct term proportional
to the gravitational breaking of $U(1)_{\rm PQ}$:
\begin{equation}
m_a = \frac{\Lambda^2_{\rm QCD}}{f} + (m_a)_{\rm gravity}~.
\end{equation}
In this case, the cosmological and astrophysical properties of axions may in
fact be quite different from those in standard invisible axion models.\cite{BS}

\subsection{Spontaneous Breaking of CP/P}

It may be possible, perhaps, to resolve the strong CP problem by imagining
that CP (or perhaps even P\cite{Mohapatra}) is a symmetry of nature,
but one which is spontaneously broken by the vacuum.  In this case, there is no QCD
vacuum angle $\theta$ and, at the Lagrangian level, Arg det $M = 0$.  Nevertheless, because CP is spontaneously broken, eventually at the loop
level one induces an effective angle $\bar\theta$.  However, one can
perhaps arrange the theory so that the resulting $\bar\theta < 10^{-10}$.\cite{spontaneous}  This, generally speaking, requires that also at
one-loop level $\bar\theta = 0$.

One can distinguish two different classes of models which try to resolve the
strong CP problem in this fashion, with their distinction being related
principally to the scale at which CP is spontaneously broken.  In the first
class of models the breaking of CP occurs at the weak scale, while in the
second class this breaking occurs at a scale close to the Planck mass.
Both types of models, however, have generic problems. In what follows, I again briefly focus on these problems.

To break CP spontaneously in the Standard Model requires having a more
complicated Higgs sector with two or more Higgs fields, each of which acquires
some complex VEV.\cite{Lee}  However, this more complicated Higgs sector,
with its CP-violating phases
\begin{equation}
\langle\Phi_i\rangle = v_i e^{i\delta_i}
\end{equation}
and with scales $v_i\sim 0~(100~{\rm GeV})$ is problematic, since it leads to
flavor changing neutral currents at an unacceptable level, unless one imposes some extra constraints.\cite{Branco}  

Although this is a troublesome feature of these kind of models, it is probably
not their worse aspect.  The spontaneous breaking of CP leads to the formation
of domains with different CP values in the early Universe.  These domains are
separated from each other by walls where considerable energy is stored.  As the
Universe cools to its present temperature, the energy density associated with
these domain walls dissipates very slowly.  For VEVs of the order of the
weak scale, the energy density which would be associated with these
domain walls in the present Universe is enormous, far exceeding the
Universe's critical density\cite{KOZ}
\begin{eqnarray}
\rho_{\rm wall} &\sim& \langle\Phi_i\rangle^3 T\sim 10^{-7} {\rm GeV}^4 \nonumber \\
& & \mbox{} \gg \rho_c  
\sim 10^{-46}~{\rm GeV}^4~.
\end{eqnarray}
Hence, it is really not tenable imagining that CP is spontaneously broken at
scales of the order of the weak scale.

Because of this cosmological problem, it has been suggested that perhaps the
spontaneous breaking of CP occurs at scales so large that inflation has not
yet taken place.  In this case, the domain wall problem disappears because
our observable Universe after inflation just occupies one of these CP domains.
Obviously having such large VEVs 
associated with spontaneous CP violation
also eliminates the FCNC problem since $v_{\rm CP-viol.} \gg 100~{\rm GeV}$.
However, the difficulty in these models resides
in transmitting the CP-violating phase generated at these high scales to the
low energy sector, so that one can actually generate the observed CP-violation
in the neutral Kaon complex.

Because
$v_{\rm CP-viol.} \gg v \sim 250~{\rm GeV}$,
it is obvious that whatever fields $\sigma_i$ are responsible for this VEV,
these fields again must all be $SU(2)\times U(1)$ singlets.  Thus, very naturally,
these kind of models have {\bf no direct} coupling of these fields to
quarks, leading to Arg det $M=0$ at tree level.  As a result, the prevalent
form of CP violation at low energy for these models occurs through the
mixing of these $SU(2)\times U(1)$ singlet fields with other Higgs fields
in the theory.  In particular, often the phases associated with the
``high scale" $SU(2)\times U(1)$ singlet VEVs are
transferred to the coupling of
triplet fields $\chi$, which can mediate directly $\Delta S=2$ transitions like
$ss\leftrightarrow dd$.\cite{Barr}  Hence, it is quite natural for models
of spontaneous CP violation at large scales to lead to superweak 
models\cite{Wolfenstein} of CP violation at low energy.  Furthermore, in these
models\cite{MMP} it is relatively easy to eliminate CP violating contributions
to Arg Det $M$ at one loop level, so that 
non-zero values of $\bar\theta$ do not
appear until 2 loops.  Obviously, we shall know relatively soon whether
superweak models for CP violation are tenable.  This may be made clear by the
next round of the $\epsilon^\prime/\epsilon$ experiments,\cite{Weinstein} but
probably will most clearly emerge from B mesons CP-violation studies at the,
soon to be operational, B factories.

I should remark that it is possible to construct models where, even though CP
is broken spontaneously at a high scale, the low energy observable CP-violation
is indistinguishable from the standard CKM model.\cite{CKM}  These types of
models were first constructed by Nelson\cite{Nelson} and Barr\cite{Barr1} and
are quite interesting.  What mitigates against them, however, is that they
are rather recondite and to make them work requires new physics at different
scales.  Basically these models have a contribution to $\bar\theta$ already at one-loop and to control this they need to invoke quite different scales.  I
have discussed in some detail the structure of these Nelson-Barr models\cite{RDP}
and do not want to repeat this
discussion here.  Suffice to say that, typically, the
one-loop contribution to Arg det $M$ is of the form
\begin{equation}
{\hbox{Arg det}} M \sim \left(\frac{M_I}{M_X}\right)^2 \times {\rm phases}
\end{equation}
where $M_X$ is a GUT scale and $M_I$ is the mass of some new fermions.  One
can guarantee $\bar\theta < 10^{-10}$ by assuming $M_I \ll M_X$.
Solving the strong CP problem in this way, by assuming a significant hierarchy
in an obscure sector of the theory, which is 
mostly decoupled from low energy physics, is clearly 
rather unsatisfactory---at least to me!

\section{CONCLUDING REMARKS}

The strong CP problem (why $\bar\theta < 10^{-10}$ rather than of $O(1)$)
is perhaps not as serious an issue as the cosmological constant problem.
After all QCD predicts $\langle T^\mu_\mu\rangle \sim 10^{-2}~{\rm GeV}^4$,
which is more than forty orders of magnitude larger than what cosmology
informs us, $\langle T_\mu^\mu\rangle < 10^{-46}~{\rm GeV}^4$!  Nevertheless,
in my view, the strong CP problem is a fairly clear indication that, at a
deep level, we really do not understand the nature of CP breaking.

I do not believe that the solution to the strong CP problem will come from
QCD itself.  Rather its solution must come from a better understanding of the
whole theory.  It is just possible that the key to the resolution of the strong
CP problem will be found when we garner a better understanding of low energy
CP-violation.  In particular, if the observed CP-violation in the Kaon system
were due to some superweak interactions,\cite{Wolfenstein} then it is possible
that $\bar\theta$ is indeed calculable and small.  We will know experimentally
soon, both from the next round of $\epsilon^\prime/\epsilon$ experiments
and from studies of B meson CP violation at the B
factories, whether low energy CP violation is described on the main by the
CKM model or  not.  If the CKM picture holds, which is my expectation, then
the strong CP problem will remain a problem.

In my view, it is likely that the solution to the strong CP problem really is
related to the existence of an effective global chiral symmetry\cite{PQ}
which makes $\bar\theta$ a dynamical parameter.  So, I am a believer in
axions and in a new dynamical scale $f$, related to the breakdown of this
overall chiral symmetry.  I am, however, less certain that the scale $f$ is in
the invisible axion range [$5\times 10^9~{\rm GeV} < f < 10^{12}~{\rm GeV}$],
although I believe it is crucially important to search for axions in this
range.  My skepticism here is connected to the perceived wiggle-room which
both alternative dynamics and cosmology provide to the determination of $f$.
Nevertheless, if $f$ is much greater than the weak scale, $f \gg v$, as it
surely is, one cannot really escape asking what is its relation to the Planck
mass, $M_{\rm P}$,

More generally, my sense is that it is very important to try to understand the
compatibility of a global chiral symmetry, like $U(1)_{\rm PQ}$ with
gravity.  Does $U(1)_{\rm PQ}$ survive gravitational effects, or not?  My
hunch is that it does and that when we will understand things better we
will find that 
the strong CP problem and the cosmological
constant problem are deeply related.  At a deep level, the solution of these
problems probably lies in string theory and supersymmetry.  In fact, in
supersymmetric theories scale and chiral transformations are naturally related, with the
dilaton and the axion both making up the scalar components of a Nambu-Goldstone
chiral superfield, and with the trace and chiral anomalies being similarly twinned.
The big question, however, is whether these musings can ever be turned into a
proper understanding of these problems!
\vspace{.3cm}

\begin{flushleft}
{\bf Acknowledgments}
\end{flushleft}
%\vspace{.3cm}

I am very grateful to Pierre Sikivie for having invited both Helen Quinn and
me to participate in this 20th anniversary celebration of 
the PQ symmetry and of axions.  Both
the meeting and the hospitality were splendid!

\end{document}